# Geometry-tunable magnetic edge contrast in $Bi_2Te_3$ Corbino nanoplates

Motahhare Mirhosseini[1,2], Swathi Kadaba[1,2], Allison Swyt[1,2], David L. Carroll[1,2*]

[1]*The Nano and Quantum Technologies Laboratory, Wake Forest University, Winston Salem NC 27105, USA*
[2]*Department of Physics, Wake Forest University, Winston Salem NC 27109, USA*

Two-dimensional topological insulators feature helical edge states that are remarkably resistant to disorder, making them appeal for energy-efficient electronics and quantum information technologies. In this study, we develop a Te-rod-templated solution growth method to create $Bi_2Te_3$ nanoplates with a Corbino geometry. The resulting few-quintuple-layer hexagonal plates are single-crystalline and contain well-defined central pores. Using optimized magnetic force microscopy, we observe clear magnetic contrast at both the inner and outer edges. The signal depends strongly on tip height and oscillation amplitude, allowing us to distinguish genuine magnetic responses from electrostatic and topographic effects. By systematically varying the pore size, we find that edge contrast increases as the distance between edges decreases, suggesting stronger coupling between the inner and outer edge channels. These findings establish a geometry-controlled platform for tuning edge-localized magnetic behavior in $Bi_2Te_3$ and open a new path to explore edge interactions in two-dimensional topological insulators
.

*Introduction*—Topological insulators (TIs) feature an insulating bulk and symmetry-protected boundary states stabilized by strong spin–orbit coupling. In the thin-film limit, where the film thickness is comparable to or smaller than the electronic coherence length, materials such as $Bi_2Te_3$ behave as two-dimensional TIs, or quantum spin Hall insulators, with one-dimensional helical edge channels that run along the sample boundaries. These edge states enable spin-momentum, locked transport and are central to proposals for dissipation-less interconnects and topologically protected qubits. [1–6]

Previous studies of $Bi_2Te_3$ have primarily examined linear edges, nanoribbons, or Hall-bar geometries, which only probe a single boundary at a time. In comparison, a Corbino geometry naturally incorporates both inner and outer edges within the same crystal, offering an ideal setting to explore edge-edge interactions. Theoretical work suggests that when the radial separation between these edges becomes small, their states can hybridize, leading to geometry-dependent coupling and distinctive energy spectra. [7,8]

Creating $Bi_2Te_3$ nanostructures with this architecture, however, remains challenging. Fabrication methods such as focused ion beam (FIB) milling or e-beam lithography can define edges with high precision, but often introduce damage, contamination, or disorders that obscure intrinsic topological features. [9,10] In contrast, solution-grown nanoplates offer atomically smooth, defect-minimized surfaces with well-defined crystal orientation, making them advantageous for studying delicate edge phenomena. [11–13]

Despite these advantages, the relationship between pore size, edge separation, and the resulting magnetic signatures has not been established experimentally. [14] Real-space evidence linking geometry to magnetic contrast would provide crucial insights into how edge modes interact and evolve-information essential for future device concepts based on edge coupling and coherent spin transport. [15]

In this work, we realize $Bi_2Te_3$ Corbino nanoplates through Tellurium (Te)-rod-guided solution growth and use magnetic force microscopy to reveal robust, edge-localized magnetic contrast that can be tuned by adjusting the geometry.

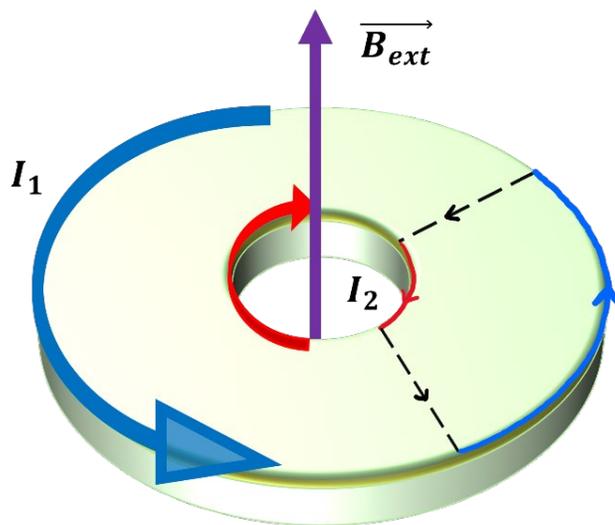

FIG. 1. Schematic of a $Bi_2Te_3$ Corbino nanoplate illustrating counterpropagating edge currents $I_1$ (outer edge) and $I_2$ (inner edge) flowing under an applied external magnetic field $\vec{B}_{ext}$ normal to the plane. The dashed arrows indicate the radial edge-edge separation.

*Synthesis of Corbino nanoplates*—$Bi_2Te_3$ nanoplates were synthesized using a solution-based route in which elemental Te first forms one-dimensional rods that act as sacrificial templates for the Corbino pores. $Bi_2Te_3$ precursors were prepared by dissolving 1.94 g of bismuth (III) nitrate and 1.33 g of sodium tellurite in 70 mL of ethylene glycol in a three-neck flask equipped with a magnetic stir bar under a slow Argon (Ar) flow. Sodium hydroxide (1.8 g) was then added to adjust the alkalinity and promote Te reduction, followed by 30 min of stirring. Polyvinylpyrrolidone (0.8 g) was introduced as a surfactant to control nucleation and stabilize the growing nanoplates, and the solution was subjected to three cycles of evacuation and Ar backfilling (5 min each) to remove dissolved gases.

During the initial heating stage, the Te precursor preferentially forms rod-like structures, which define the eventual pore position within each nanoplate. As the reaction temperature approaches 150 °C, Bi and Te co-nucleate and grow laterally around these Te rods, producing hexagonal $Bi_2Te_3$ plates with the Te rod at the center. Further heating toward 160 °C fully crystallizes the $Bi_2Te_3$, while the Te template gradually dissolves or detaches from the surrounding plate, leaving a clean central pore.

The pore size was tuned by controlling the Te-rod thickness and growth conditions. In practice, increasing the relative Te content, adjusting NaOH concentration, and modifying the early-stage temperature ramp all lead to thicker Te rods and therefore larger pores, whereas lower Te loading and milder basicity favor thinner rods and smaller pores. By systematically varying these parameters, we obtained nanoplates with reproducible pore diameters across a broad range while maintaining good crystallinity and mechanical stability.

For comparison, we also explored a post-annealing approach, in which solid $Bi_2Te_3$ plates were first synthesized without pores and then subjected to a secondary thermal treatment to induce void formation. In this method, pores formed irregularly and often coalesced, and the additional thermal cycling frequently produced cracked or partially fractured plates. In contrast, the Te-rod-guided growth described above generated well-defined, centrally located pores and mechanically robust $Bi_2Te_3$ Corbino nanoplates suitable for systematic edge studies.

*Structural Characterization*—Transmission electron microscopy (TEM) and selected-area electron diffraction (SAED) were used to examine crystal structure across pore edge regions (Figure S2) . [16] TEM images of Te-rod-grown nanoplates show uniform hexagonal geometry with sharp edges and a well-defined central pore, while SAED patterns from distinct regions exhibit essentially identical diffraction spots consistent with a single-crystalline rhombohedral R-3m $Bi_2Te_3$ lattice. [16,17]

Scanning TEM (STEM) with high-angle annular dark-field (HAADF) imaging and energy-dispersive X-ray spectroscopy (EDS) confirms that Bi and Te are distributed homogeneously across the nanoplate, including near the pore, indicating uniform stoichiometry despite the complex geometry (Figure S3). Atomic force microscopy (AFM) height maps reveal flat surfaces and a uniform thickness plateau around 5 nm, with sharp step edges at the boundaries, consistent with a few-quintuple-layer two-dimensional TI regime.[1–3]

*Magnetic Force Microscopy*—Magnetic force microscopy measurements were optimized by systematically tuning both the oscillation amplitude and the lift height to maximize genuine edge-localized magnetic contrast while minimizing topographic, electrostatic, and capillary artifacts. [15,20,21]

The oscillation amplitude was first varied at fixed lift height, as shown in Figure 3(a–e), to determine the amplitude range that yields the strongest, spatially confined phase shifts at the inner and outer edges. In this configuration, phase images acquired at very small amplitudes are dominated by topographic and electrostatic noise, while an intermediate amplitude window of approximately 10–18 nm produces pronounced and well-resolved edge contrast for both inner and outer edges, consistent with prior optimization studies of MFM sensitivity versus drive amplitude. Beyond roughly 20 nm, the tip trajectory spans both edge and non-edge regions within a single oscillation cycle, which averages the local force gradient, broadens the apparent edge width, and reduces the peak phase shift, in line with reports that excessive drive amplitudes degrade lateral resolution despite increased force sensitivity.

The amplitude dependence of the inner and outer edge phase shifts extracted from these images is summarized in Figure 3(f), which reveals a maximum in both signals for amplitudes in the 10–18 nm range. On this basis, an oscillation amplitude of 17.4 nm was selected for all subsequent measurements as a compromise that maximizes edge contrast while avoiding tip–sample instabilities or spurious interactions associated with larger oscillation envelopes.

With the oscillation amplitude fixed at 17.4 nm, the lift height was then incremented from 0 to 100 nm in 20 nm steps, as illustrated in Figure 3(g–l), to decouple long-range magnetic forces from short-range van der Waals, capillary, and electrostatic contributions. At the smallest lift heights (0–10 nm equivalent), strong non-magnetic interactions lead to significant phase noise and image distortions, consistent with general MFM practice that warns against too small a separation where topography artifacts dominate despite the nominal gain in sensitivity. As the lift height increases, these short-range forces decay rapidly while the more slowly varying magnetic stray fields remain detectable, yielding clearer edge-localized phase contrast over an intermediate lift-height window, before the signal gradually weakens at very large separations as the magnetic force gradient falls below the detection limit.

In this optimized regime, the phase signal associated with the outer edge persists to higher lift heights than that of the inner edge, reflecting differences in the local stray-field profiles and effective magnetic charge distributions, as observed in previous high-resolution MFM studies of magnetic domains, edges, and skyrmionics textures. The final operating conditions, oscillation amplitude of 17.4 nm and lift height chosen within the range where the edge contrast is maximized and non-magnetic artifacts are suppressed, follow established MFM optimization strategies and enable quantitative mapping of nanoscale magnetic structures in analogy to earlier work on magnetic domains, domain walls, and chiral spin textures. [15,21,22]

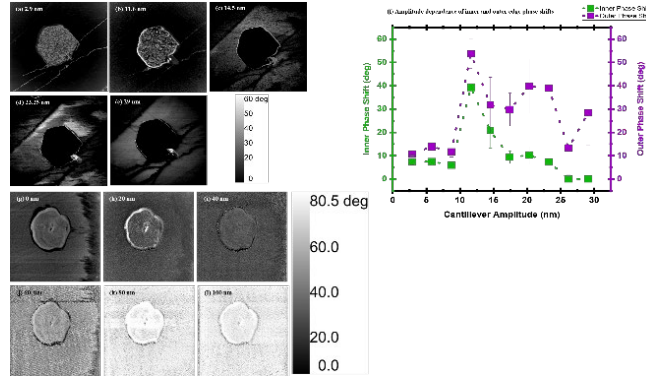

FIG. 2. MFM optimization and edge localization. (a–e) MFM phase images acquired at a fixed lift height for oscillation amplitudes of 2.9, 11.6, 14.5, 23.25, and 29 nm, respectively, illustrating the emergence and subsequent broadening of edge contrast with increasing amplitude. (f) Amplitude dependence of the inner (green) and outer (magenta) edge phase shifts, highlighting an optimal amplitude window of about 10–18 nm. (g–l) MFM phase images acquired at a fixed oscillation amplitude of 17.4 nm for lift heights of 0, 20, 40, 60, 80, and 100 nm, demonstrating the optimization of lift height for resolving inner and outer edge magnetic contrast.

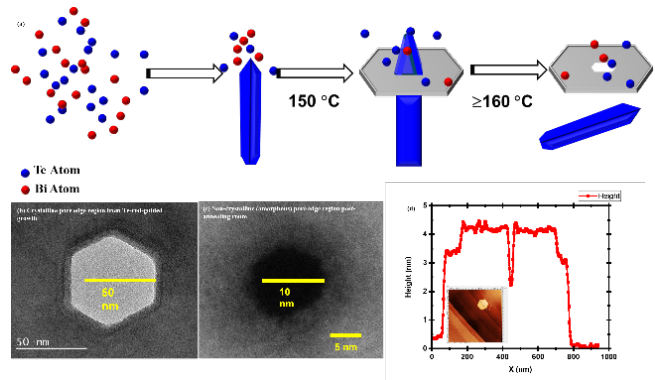

FIG. 3. (a) Schematic of the Te-rod–guided solution growth process in which Bi and Te atoms nucleate around a sacrificial Te rod near 150 °C and form a Corbino nanoplate as the rod dissolves above 160 °C. (b) TEM image of a Te-rod–grown $Bi_2Te_3$ nanoplate showing a well-defined hexagonal central pore with ~50 nm diameter. (c) TEM image of a nanoplate grown by a failed post-annealing route, exhibiting an irregular pore and strong structural distortion. (d) AFM map and corresponding line profile confirming a flat surface and uniform thickness of ~5 nm.

*Edge-localized magnetic contrast*—Under optimized MFM conditions, the MFM phase signal is strongly localized along the inner and outer circumferences of the $Bi_2Te_3$ Corbino nanoplates, while the interior bulk region exhibits significantly weaker contrast.[9,10] The emergence of edge contrast only beyond a threshold lift height, together with its suppression at excessively large oscillation amplitudes, supports a magnetic rather than purely electrostatic origin.[6,8,11] The persistence of outer-edge contrast to larger lift heights suggests a higher density or stronger response of outer-edge states compared to inner-edge states.

A representative MFM phase image shows a bright or dark ring signal closely following both the inner and outer edges, consistent with edge-localized magnetic field gradients. [23,24] The spatial correlation between phase features and the structural edges observed in TEM and AFM confirms that the signal originates from the Corbino boundaries and not from random inhomogeneities in the bulk.

*Geometry-dependent edge contrast*—To probe edge-edge coupling and geometry dependence, nanoplates with different pore sizes and thus different inner-outer edge separations (w) were examined.[4,12,13] Larger pores correspond to smaller radial gaps and longer inner and outer circumferences, providing more edge length and stronger expected overlap of edge-state wavefunctions.[14,15] MFM measurements show that the absolute magnitude of edge phase contrast increases monotonically as w decreases, for both inner and outer edges.[14,26]

Quantitatively, the average phase shift along each edge, extracted from line profiles or azimuthal averages, grows as the gap narrows, indicating a geometry-tunable enhancement of edge-localized magnetic field gradients.[4,12,13] This monotonic dependence is consistent with theoretical expectations for hybridizing inner and outer edge channels in a Corbino geometry, where reduced separation strengthens edge-edge coupling and increases the effective local response.[4,12,13]

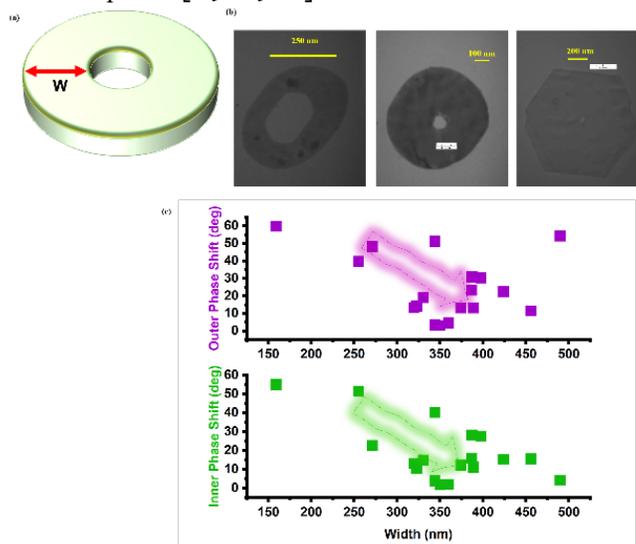

FIG. 4. Geometric dependence of edge-localized magnetic contrast. (a) Schematic of an annular platelet defining the ring width "w". (b) Representative TEM images of platelets with different widths. (c) Outer (top, magenta) and inner (bottom, green) MFM phase shift as a function of ring width, showing a systematic decrease in edge phase contrast with increasing width.

*Discussion*—The combination of structural and magnetic measurements presents a self-consistent picture of edge-localized, geometry-tunable contrast in $Bi_2Te_3$ Corbino nanoplates. High-quality TEM, SAED, STEM-EDS, and AFM data demonstrate that the nanoplates are single-crystalline, chemically homogeneous, and uniformly thin, reducing the likelihood that the observed MFM signals arise from gross defects or compositional inhomogeneities at the edges.[1,12–18] Instead, the strong spatial confinement of the signal to inner and outer boundaries, and its systematic increase with decreasing edge-edge separation, point toward edge-specific phenomena sensitive to geometry.[4,12,13]

Several microscopic mechanisms could contribute to the observed magnetic contrast. One candidate is helical edge states, whose spin structure and current distribution might generate edge-localized magnetic fields.[1,3,22–26] Alternatively, strain and defect-induced magnetism have been reported in topological materials and could also contribute to edge-localized magnetic responses.[7,27] At the same time, topological insulators are robust against moderate disorder, so strain is expected to tune band dispersion and edge-state hybridization without necessarily destroying the nontrivial topology, consistent with a scenario where geometry and strain jointly modulate edge-edge coupling within a 2D-TI phase.[22,23,28,29]

Disentangling among these possibilities will require temperature- and field-dependent MFM, as well as comparison with non-topological reference materials. If the edge contrast remains robust over a wide temperature range and shows characteristic responses to applied magnetic fields consistent with helical or chiral edge channels, it would support a topological
origin.[1,4,17,30] By contrast, strong suppression at elevated temperatures or similar behavior in non-topological reference materials would point toward defect- or strain-dominated magnetism.[44,45] Such studies, combined with approaches used to separate mechanisms in magnetic TIs and axion-insulator candidates, will help clarify the microscopic origin.[34,40,46–48] However, we suggest such an effect would be small when compared to the effect of changing "w" as theoretically predicted.[14]

*Conclusions*—Bi$_2$Te$_3$ Corbino nanoplates synthesized via a Te-rod-templated solution route form a robust platform for real-space studies of edge physics in two-dimensional topological insulators. [16,27,29,30,49] Magnetic force microscopy, optimized in lift height and oscillation amplitude, reveals stable magnetic contrast localized at both inner and outer edges and negligible signal in the bulk interior. [20–22,25] Systematic variation of pore size shows that the magnitude of edge contrast increases monotonically as the edge-edge separation decreases, demonstrating geometry-tunable edge-localized magnetic signatures consistent with enhanced inner-outer edge coupling. [14,19,26]

These findings establish a route to control and probe topological edge interactions through simple geometric design and lay the groundwork for future experiments that exploit Corbino nanoplates as building blocks for edge-engineered quantum devices. [45,50]

*Acknowledgments*—This research was supported by Wake Forest University Graduate School of Arts and Sciences. All Authors thank Dr. Chaochao Dun for valuable contributions to the selected-area electron diffraction measurements performed at the Molecular Foundry, Lawrence Berkeley National Laboratory. The authors also thank Dr. Chris Wrinkler at NC State for assistance with HRTEM work at the Analytical Instrumentation Facility (AIF) at North Carolina State University. The AIF is a member of the North Carolina Research Triangle Nanotechnology Network (RTNN), a site in the National Nanotechnology Coordinated Infrastructure (NNCI)

*Data availability*—The data that supports the findings of this article are not publicly available. The data are available from the authors upon reasonable request.